\def \SAIT #1 #2 {{\em Mem.\ Soc.\ Astron.\ It.\/} {\bf #1}, #2}
\def \MESS #1 #2 {{\em The Messenger\/} {\bf #1}, #2}
\def \ASTRNACH #1 #2 {{\em Astron. Nach.\/} {\bf #1}, #2}
\def \AAP #1 #2 {{\em Astron. Astrophys.\/} {\bf #1}, #2}
\def \AAL #1 #2 {{\em Astron. Astrophys. Lett.\/} {\bf #1}, L#2}
\def \AAR #1 #2 {{\em Astron. Astrophys. Rev.\/} {\bf #1}, #2}
\def \AAS #1 #2 {{\em Astron. Astrophys. Suppl. Ser.\/} {\bf #1}, #2}
\def \AJ #1 #2 {{\em Astron. J.\/} {\bf #1}, #2}
\def \ANNREV #1 #2 {{\em Ann. Rev. Astron. Astrophys.\/} {\bf #1}, #2}
\def \APJ #1 #2 {{\em Astrophys. J.\/} {\bf #1}, #2}
\def \APJL #1 #2 {{\em Astrophys. J. Lett.\/} {\bf #1}, L#2}
\def \APJS #1 #2 {{\em Astrophys. J. Suppl.\/} {\bf #1}, #2}
\def \APSS #1 #2 {{\em Astrophys. Space Sci.\/} {\bf #1}, #2}
\def \ASR #1 #2 {{\em Adv. Space Res.\/} {\bf #1}, #2}
\def \BAIC #1 #2 {{\em Bull. Astron. Inst. Czechosl.\/} {\bf #1}, #2}
\def \JSQRT #1 #2 {{\em J. Quant. Spectrosc. Radiat. Transfer\/} {\bf #1}, #2}
\def \MN #1 #2 {{\em Mon. Not. R. Astr. Soc.\/} {\bf #1}, #2}
\def \MEM #1 #2 {{\em Mem. R. Astr. Soc.\/} {\bf #1}, #2}
\def \PRL #1 #2 {{\em Phys. Rev. Lett.\/} {\bf #1}, #2}
\def \PASJ #1 #2 {{\em Publ. Astron. Soc. Japan\/} {\bf #1}, #2}
\def \PASP #1 #2 {{\em Publ. Astr. Soc. Pacific\/} {\bf #1}, #2}
\def \NAT #1 #2 {{\em Nature\/} {\bf #1}, #2}
\def \be {\begin{equation}}
\def \ee {\end{equation}}

\documentstyle{memsait}
\input epsf.sty
\begin{opening}
\title{ON THE NATURE OF SOFT GAMMA-RAY REPEATERS}
\author{BING ZHANG}
\institute{Astronomy \& Astrophysics Department, 
Pennsylvania State University,  USA}
\date{} 
\end{opening}

\begin{document}

\oddpagefooter{}{}{} 
\evenpagefooter{}{}{} 
\ 
\bigskip

\begin{abstract}
Soft gamma-ray repeaters (SGRs) and Anomalous X-ray pulsars (AXPs) are 
generally accepted to be magnetars. Recently, Zhang, Xu \& Qiao (2000, 
ApJ, 545, L127) proposed an alternative viewpoint about the nature of
the SGRs (and AXPs). In this picture, SGR bursts are attributed to the
impacts of some comet-like objects with the central bare strange
star. Here I briefly review this model, and confront it with the
detailed observations of SGRs/AXPs. A comparison to the magnetar 
model is also presented. Some theoretical issues concerning
the nature of the SGRs (and AXPs) are discussed. 
\end{abstract}

\section{An impacting model for SGRs}
Soft gamma-ray repeaters (SGRs) and anomalous X-ray pulsars (AXPs) are 
distinct among other astrophysical objects in several aspects. 1. They
are peculiar among the categories they were originally belong to: SGRs 
are anomalous among the normal gamma-ray bursts, and AXPs are
anomalous among the normal X-ray pulsars. 2. These two groups of
objects emerging from two distinct astrophysical branches share many
common properties (they are also different in some aspects,
though). 3. The nature of these objects, as 
discussed below, is very likely to be exotic, either a neutron star
with ultra-high magnetic fields (a magnetar), or a ``neutron star''
with exotic interior (a strange star or a hybrid star), or even a
combination of the both. Any successful theory should address both the
similarities and the discrepancies between SGRs and AXPs.

A well-accepted model for SGRs/AXPs is the magnetar model
(see Thompson 2001 for a review). A ``nurture'' model for the AXPs,
which invokes accretion from a fossil accretion disk, is also
prevailing (e.g. Chatterjee et al. 2000), but no attempt was made to
interpret SGR bursts within this model. Recently, we (Zhang, Xu \& Qiao 2000,
hereafter ZXQ00) proposed a ``nurture + nature'' model for SGRs. There 
are three independent assumptions in this model: 1. plenty of comets
exist in the ``Oort Cloud'' of the massive progenitor of the SGR host, 
and they occasionally impact the SGR host when the host passes through 
the Oort Cloud; 2. a fossil accretion disk is formed after the
supernova explosion, and accretion from this disk onto the surface
of the SGR host is responsible for the long-term spindown behavior of
the SGR; 3. Stars composed with strange quark matter, i.e., strange
stars, exist in nature, and they are the SGR hosts. A detailed
description about how these assumptions can account for the SGR
phenomenology is presented in ZXQ00. Here I only highlight two
relevant issues: 1. The ``nurture'' part of the model, i.e., the
comet cloud, is more essential for the model, which gives an
alternative mechanism (other than magnetic fields) to power the SGR
bursts. The ``nature'' part of the model, i.e., strange stars, may be
not definitely necessary (see more in \S 3), although their existence
makes interpretations easier. 2. An important expectation of this
model is the quasi-periodicity of the SGR bursting behavior due to the
orbital motion of many comets circulating the SGR host. This results
in a direct prediction, i.e., SGR 1900+14 will become active again in
the year 2004-2005.

\section{Confronting models with observations}

\subsection{Spindown behavior}
The prominent properties of SGRs/AXPs are their long periods and high
spindown rates comparing with normal radio pulsars. In the
conventional $P-\dot P$ diagram, they form a separate island in the
upper right corner. Periods are clustering within a narrow range,
i.e., 5-12 s. No spin-up is ever observed from these objects. The
spindown behavior is usually steady, but not always. Increases of the
spindown rate have been observed in two SGRs, which are not necessarily
related to the bursting behavior. With $P$ and $\dot P$, one can
define the characteristic age $\tau \sim P/2\dot P$, which is typically
$10^3-10^4$ year, and the dipolar surface magnetic field $B_s =
6.4\times 10^{19}{\rm G}(P\dot P)^{1/2}$, which is typically $10^{14}
-10^{15}$G. But these estimates are not fully reliable due to the
variable $\dot P$. Glitches might have been observed from some
SGRs/AXPs, but with diverse characteristics. More specifically, the
glitch accompanied with the August 27 event of SGR 1900+14 has an
opposite sign to those observed in normal pulsars.

The magnetar model gives a straightforward
interpretation to the long $P$ and high $\dot P$ of the SGRs/AXPs. 
Electromagnetic dipolar spindown with the magnetic fields inferred
from the spindown data can naturally interpret the steady spindown,
and the inclusion of some other processes, such as an Alfven-wave
driven particle outflow or free precession of the star due to magnetic 
distortion, can interpret the viable spindown behavior,
including the ``negative'' glitch (e.g. Thompson et al. 2000). A
question is how substential increase of the wind luminosity could be
not related to the bursts.

An alternative model is to attribute the peculiar spindown behavior to
accretion from a fossil disk formed after the supernova explosion,
which is adopted in the model of ZXQ00. In this model, the spindown
behavior of the SGRs/AXPs is also well-interpreted (Chatterjee et
al. 2000). The irregular spindown observed in some objects could be
due to slightly variation of the accretion rate.
It has been argued that long-term steady spindown in some AXPs could
be an evidence against the accretion scenario. This criticism is not
robust since at least one accretion system has been steadily spinning
down for 15 years (Baykal et al. 2000). The ``negative'' glitch
accompanied with the August 27 event may be interpreted in the ZXQ00 
model by assuming that the torque exerted by the impact
happened to be opposite to the angular momentum of the star.

\subsection{Counterparts}
No counterpart of SGRs/AXPs in other wavebands has been firmly
detected until the recent report by Hulleman et al. (2000), who
discovered an optical counterpart for the AXP 4U 0142+61. The authors
claimed that optical emission is too faint to admit a large accretion
disk, but may be consistent with magnetospheric emission from a
magnetar, although no optical emission model for magnetars is
known. It is unclear whether this observation is indeed a support to
the magnetar model, but it definitely constraints the accretion
models. Two caveats ought to be kept in mind: 1. Invoking beaming
effect or truncating inner part of the fossil disk may still revive
the accretion scenario (Perna \& Hernquist 2000); 2. This particular
AXP has no supernova 
association, which may be intrinsically different (e.g. magnetic white
dwarfs) from other AXPs and SGRs.

If the fossil-disk accretion scenario has to be abandoned, the
impacting model for SGRs (ZXQ00) ought to be modified to interpret the 
large spindown rate. 
Besides the magnetar idea, an alternative approach is to appeal to the
peculiar properties of a strange star or a hybrid star as the central
object. If the star can excite substential long-term wind throughout 
its life, the SGRs/AXPs can keep spinning down rapidly with
normal field strength below critical. At present, any operative
proposal is lacking.

No SGR/AXP is firmly detected to have pulsed radio emission. Within
the magnetar model, this very likely requires that photon splitting
operate for both polarization modes in the magnetar environment which
completely suppresses pair production (see Zhang \& Harding, 2001, and
references therein). 
In the accretion scenario, lack of radio emission is straightforward
since accretion prohibits any outflow. That a high magnetic field
radio pulsar with similar spin parameters to the AXP 1E 2259+586 
lacks strong X-ray emission is a bit problematic
for the magnetar model. Something special (much stronger multipole
field or different magnetic field origin) has to be assumed at least
for 1E 2259+586. The accretion scenario can explain these
complications naturally.

\subsection{Environments}
All SGRs and some AXPs have supernova remnant (SNR) associations. If
these associations are real, the ages of the SNRs are found to be
usually (an order of magnitude) longer than the ``characteristic age''
$\tau$ measured from the timing parameters. In some cases (e.g. 1E
2259+586), the discrepancy is the other way round. Assuming SNR
associations, the inferred proper motion velocities of the SGRs are
generally much higher than those of the normal pulsars. AXPs, however,
are almost sitting right in the center of their SNRs, inferring much
slower proper motions. SGRs/AXPs are further found to be associated
with some dense environments (Rothschild et al. 2001).  Two SGRs are
associated with two compact massive star clusters, respectively.

In the magnetar model, the shorter characteristic ages with respect to
the SNR ages (which may be more reliable) are compatiable with the
existence of the winds, but the dipolar field strengths are lowered
and it is unclear whether the reduced field strengths are still enough
to power both the quiescent emission and the bursts. The longer
characteristic age for 1E 2259+586 is even problematic, which requires
the source to have much larger $\dot P$ in the past, while
observations show that this AXP is among those with most steady
spindown. The existence of winds makes the things worse. In the
accretion scenario, all these are no longer a problem, since the
characteristic age is no longer meaningful. There is no explanation
within the magnetar model about the large discrepancy between the
proper motion velocities of the SGRs and the AXPs. In the picture of
ZXQ00, AXPs are assumed to be still neutron stars, while SGRs are
assumed to be strange stars, which might have received an additional
``kick'' during the supernova explosions. It is also not straightforward
why a magnetar should be accompanied with a dense
environment and a compact cluster, although their existence do not
directly expel the magnetar model. On the other hand, the dense
environment favors the formation of the fossil accretion disk, and the
compact clusters are ideal places where plenty of comet-like objects
required by the impacting model (ZXQ00) may be available.

\subsection{Quiescent emission}
Both SGRs and AXPs have quiescent pulsed emission with luminosities of 
$L_x \sim 10^{35}-10^{36} {\rm ergs~ s^{-1}}$, well in excess of the
spindown luminosity $L_{\rm sd}=4\pi^2 I\dot P P^{-3}$. AXPs usually
have a blackbody + steep power-law spectrum. SGRs usually have a
slightly flatter power-law, and a blackbody component is also observable
sometimes. Recently, Kulkarni et al. (2001) discovered that SGR
0525-66, an SGR which was active 20 years ago and has been inactive 
since then, has a similar steep power-law as those of the AXPs.

Quiescent emission is interpreted as magnetic field decay or enhanced
neutron star cooling within the magnetar model. The high pulsation
amplitudes of AXPs may contraint some subtypes of these models. 
The magnetic field decaying luminosities should be positively
correlated to the field strength in the decaying model, but such a
correlation is not seen from the data, e.g., 1E 2259+586
has the weakest 
dipolar magnetic field, but a much higher quiescent luminosity than
some other SGRs/AXPs.  
In the accretion scenario, no direct
correlation between the X-ray luminosity and $\dot P$ is expected due
to the unknown field strength of the star, and 
high pulse amplitudes are natural due to the accretion column in the
polar cap region. However, if the optical counterparts from more AXPs
rule out the accretion scenario, some other energy sources have to be
incorporated if SGRs/AXPs are not magnetars. A possible source is
quark deconfinement (i.e. phase conversion) energy from the central
star, which could be either a strange star covered by a crust (Cheng
\& Dai 1998) or a slowly-contracting ``hybrid'' star (Dar \& de
R\'ujula 2000). No reasonable operative mechanism has been fully
proposed so far. 
The steep power-law spectrum of SGR 0525-66 strongly supports the
magnetar model, in which AXPs are inactive SGRs so that a SGR should
resemble an AXP after it enters the quiescent phase for a while. It is 
unclear whether this feature could be incorporated in the impacting
model. There are two other arguments against the accretion models
(Thompson et al. 2000): 1. the increase in persistent $L_x$ after a
burst is inconsistent with a constant spin-down torque; 2. it is
impossible to recover the quiecent emission within one day after the
August 27 giant flare since the disk may have been evaporized. These
criticisms are valid if bursts are also powered by the disk accretion, 
but may be invalid for the impacting model (ZXQ00) since the comet impacts are 
an additional energy source and that they 
usually occur in the off-polar-cap region.

\subsection{Burst characteristics}
A prominant feature is that SGR bursts are super-Eddington with 
luminosities $L_b \sim 10^{38}-10^{42}{\rm ergs~s^{-1}}$. The fluence
distribution is a power law and the waiting time distribution is
lognormal. Two giant flares share very similar properties and very
detailed information has been collected (see more in Kouveliotou
2001). 

In superstrong magnetic fields, the Compton cross section for E-mode
photons is strongly suppressed, and the enhanced magnetic Eddington
limit could be much higher than the conventional Eddington limit. SGR
repeating bursts are interpreted as crust crackings, thus a power law
fluence distribution is natural since it has been observed in
earthquakes. Giant flares are due to large scale field reconnection,
and detailed modeling has been done which can well interpret the
August 27 event. Successfully interpreting the SGR burst phenomenology
is a key strength of the magnetar model. In the impacting model (ZXQ00),
the SGR host is a bare strange star, which may not subject to the
Eddington limit at all. The power law fluence distribution can be
attributed to a power law distribution of the comet mass (Salpeter's
mass function). The bursting spectra and light curve have been
studied by some other authors within the neutron star impacting model. More
detailed work needs to be done to meet the recent observations.

\section{Some theoretical issues}

$\bullet$ Conciseness

In terms of conciseness, the magnetar model is more elegant in that
only one assumption (strong fields, i.e. nature) is
made. The impacting model is more messy which requires three
independent assumptions (\S 1). However, the special environments of
SGRs may indeed require at least two assumptions (nature + nurture),
unless one can justify the connection between the nature of the star
and the special environments.
\vskip 5pt

\leftline{$\bullet$ Is a strange star necessary?}
In ZXQ00, the necessity of invoking the strange star hypothesis is
mainly to interpret the super-Eddington luminosities. Katz (1996) claims that
super-Eddington luminosity is achievable in neutron stars if the
energy transfer is through magnetic fields. In this
sense, the strange star may be not definitely necessary, but one needs 
to confront the baryon contamination and the large proper motion
problems which may be solvable in the strange star picture (ZXQ00).
However, if the fossil-disk scenario is eventually ruled out, 
the strange star hypothesis may be also
necessary to interpret the large spindown and quiescent emission,
although any operative model has yet been proposed. 
\vskip 5pt

\leftline{$\bullet$ Is the strange star necessary to be bare?}
The ``bare'' strange star invoked by ZXQ00 could exist as long as:
1. the fallback materials in the initial phase of the supernova explosion 
are not accreted onto the surface (but are ``propelled'' away), and
2. the accretion luminosity in the ``tracking phase'' is below $\sim
4\times 10^{36}{\rm ergs~s^{-1}}$. Although the second condition is
satisfied for SGRs/AXPs, the case for the first condition is unclear. 
One thing should be commented is that even if the strange star have a
crust, the merit of super-Eddington luminosity may still pertain in
the impact picture,
since the impact may punch a hole all the way down to the quark core
and the hole will allow super-Eddington emission (Z. G. Dai, personal
communication).  
However, if the bursts are not powered by impacts, but by some
internal processes (Cheng \& Dai 1998), the Eddington limit may
not be avoided.
\vskip 5pt

\leftline{$\bullet$ Magnetar, strange star, or strange magnetar?}
The magnetar hypothesis is not less exotic than the strange star
hypothesis. The possibility of sustaining superstrong magnetic fields
by a neutron star was not justified before. recently, P\'erez Martinez et
al. (2000) pointed out that under the superstrong magnetic fields
conjectured in the magnetar model, the neutron gas pressure in the
equatoral direction is too small to balance the magnetic pressure, so
that the star will endure a transverse collapse to form a strange star 
or a hybrid star with a normal magnetic field. Thus magnetars can be only 
formed temporally, (i.e. those conjectured to power the cosmic
Gamma-ray bursts may exist), but there is no stable magnetars to power
SGRs/AXPs. 
This effect, however, provides a natural mechanism (otherwise may be
questionable) to form strange stars. If such a criticism is valid, 
some other mechanisms such as the impact picture conjectured by ZXQ00 may be
responsible for the SGR bursts. On the other hand, Xu \& Busse (2000)
argued that strange stars, during their birth, may also undergo
vigorous convection and dynamo actions to achieve superstrong magnetic 
fields. Since the star is completely composed of charged quarks, it
may be free of the transverse collapse discussed by P\'erez Martinez et
al. Thus, magnetars, if they do exist in nature, may have to be strange
magnetars. 
\vskip 5pt
To conclude, the magnetar model is successful in interpreting most of
the SGR/AXP phenomenology, but some issues (including their existence)
are not fully satisfactory. The impacting strange star model (ZXQ00)
may also interpret most of the observations, but more work needs to be
done to compare with the detailed data, especially if the fossil-disk
accretion scenario is ruled out. Whether SGR 1900+14 becomes active in
2004-2005 may be a key criterion to differentiate between the two scenarios.

\acknowledgements
I thank NASA NAG5-9192 and NAG5-9193 for support and
R. X. Xu and G. J. Qiao for stimulative collaborations.  



\end{document}